# CATEGORIZATION OF TABLAS BY WAVELET ANALYSIS

Anirban Patranabis[1], Kaushik Banerjee[1], Vishal Midya[2], Shankha Sanyal[1], Archi Banerjee[1], Ranjan Sengupta[1] and Dipak Ghosh[1]

[1]Sir C V Raman Centre for Physics and Music, Jadavpur University, Kolkata 700032, India
[2]Indian Statistical Institute, Kolkata, India

**Abstract**

Tabla, a percussion instrument, mainly used to accompany vocalists, instrumentalists and dancers in every style of music from classical to light in India, mainly used for keeping rhythm. This percussion instrument consists of two drums played by two hands, structurally different and produces different harmonic sounds. Earlier work has done labeling tabla strokes from real time performances by testing neural networks and tree based classification methods. The current work extends previous work by C. V. Raman and S. Kumar in 1920 on spectrum modeling of tabla strokes. In this paper we have studied spectral characteristics (by wavelet analysis by sub band coding method and using torrence wavelet tool) of nine strokes from each of five tablas using Wavelet transform. Wavelet analysis is now a common tool for analyzing localized variations of power within a time series and to find the frequency distribution in time–frequency space. Statistically, we will look into the patterns depicted by harmonics of different sub bands and the tablas. Distribution of dominant frequencies at different sub-band of stroke signals, distribution of power and behavior of harmonics are the important features, leads to categorization of tabla.

**Introduction**

The ubiquity of tabla in the Indian Subcontinent is without dispute. Its variety of tonal colors gives it a flexibility seldom matched by other percussion instruments. Among the percussion instruments, tablas plays an important role in accompanying vocalists, instrumentalists and dancers in every style of music from classical to light in India. . Tabla is comprised of a pair of drums, a treble drum, referred to as the tabla or *dayan*, and the bass drum called the *bayan*. The *bayan* is made of copper, brass, iron or terracotta (clay). The right-hand *dayan* is a tapering cylinder carved from a block of dense wood. Each of the drums is covered with goatskin. Undoubtedly the most striking characteristic of the *tabla* is the large black spot on each of the playing surfaces called the *syahi*. These black spots are a mixture of gum, soot, and iron filings. This is applied to the center of the *dayan*, and off-center on the *bayan*. Their function is to create the bell-like timbre that is characteristic of the instrument. The complexity of its construction accounts for its versatility. This complexity reaches such a degree that only trained craftsmen can create a tabla. When a membrane stretched over a resonating body is struck, there is generally no clear sense of pitch because the sound produced is rich in inharmonic overtones. In Parag Chordia's paper (Parag chordia, CCRMA) he described a system that segments and labels tabla strokes from real performances. The work was done with large and diverse database includes the methodology of testing neural networks and tree based classification methods. Tabla strokes are typically inharmonic in nature but strongly pitched resonant strokes (C V Raman, 1934). The sounds of most drums are characterized by strongly inharmonic spectra; however tablas, especially the dayan are an exception. This was pointed out in 1920 by C. V. Raman and S. Kumar. C V Raman further refined the study in a later paper in 1934. The classical model put forth by Raman represents the sound of tabla-dayan, as having a spectrum consisting of five harmonics; these are the fundamental with its four overtones. Thereafter several theoretical and experimental studies were held on the dynamics of the instrument (R N Ghosh, 1922; K N Rao,1938; B S ramakrishna,1957; T Sarojini et. al., 1958; B M Banerjee et.al., 1991). In this paper we studied the harmonic analysis and distribution of power within the time series of tabla strokes.

The two drums of the tabla produce many different timbres. Many of these sounds have been named, forming a vocabulary of timbres. The naming of strokes has facilitated the development and transmission of a sophisticated solo repertoire. In addition to the rhythmic complexity of tabla music, it is its timbral beauty and diversity that distinguish it from other percussion instruments (C V Raman and S Kumar, 1920). In this paper we have considered five different tablas and nine single strokes from each table. Stroke 'Ta/Na' executes by lightly pressing the ring finger down in order to mute the sound while index finger strikes the edge. Stroke 'Ti' executes by striking the dayan on the 2nd circle with the index finger and by keeping the finger on that position causes more damping but after striking if the index finger release quickly to give an open tone it produces 'Teen'. Stroke 'Ghe' executes by striking the bayan with middle and index finger keeping the wrist on the membrane but after striking if released quickly it produces 'Ge'. Stroke 'Thun' executes by striking on the center circle of dayan with index, middle, ring and little fingers together and by quickly releasing. Stroke 'Tu' executes by striking at the corner of center circle of dayan with index finger only and immediately after striking finger will lift. Stroke 'Te' executes by striking the dayan with middle and ring finger at the center of the circle. Stroke 'Re' executes by striking the dayan with index finger at the center of the circle and by keeping the finger on that position causes more damping.

## Experimental Setup

All the strokes were played by eminent tabla players and the sound was recorded in a noise free acoustic room. Membrane of tabla 1, 2 and 3 have diameter 5", tabla 4 has a diameter 5.5" and tabla 5 has a diameter 6". Each of these sound signals was digitized with sample rate of 44.1 kHz, 16 bit resolution and in a mono channel. We then undertook iterated filter bank by sub-band coding for analyzing localized variations of amplitude within a time series and to find the frequency distribution in time–frequency space. We also used Torrence wavelet tool to observe distribution of power and behavior of harmonics. Timbre analysis was done by Long Term Average spectra (LTAS) of each signal.

## Wavelet analysis

The whole purpose of wavelet analysis is to see the time-scale (frequency) distribution, i.e. how the power changes over time and to find the frequency distribution in time–frequency space. The idea behind these time-frequency joint representations is to cut the signal of interest into several parts and then analyze the parts separately. It is clear that analyzing a signal this way will give more information about the when and where of different frequency components present (C Valens, 1999-2004). The continuous wavelet transform or CWT. is written as:

$$\gamma(s, \tau) = \int f(t) \psi^*_{s,\tau}(t) dt$$

Where * denotes complex conjugation. This equation shows how a function f (t) is decomposed into a set of basic functions $\psi_{s,\tau}(t)$, called the wavelets. The variables s and τ are scale and translation, are the new dimensions after the wavelet transform. If we expand the wavelet transform into the Taylor series at t = 0 until order n (let = 0 for simplicity) we get

$$\gamma(s,0) = \frac{1}{\sqrt{s}} \left[ f(0)M_0 s + \frac{f^{(1)}(0)}{1!} M_1 s^2 + \frac{f^{(2)}(0)}{2!} M_2 s^3 + \ldots + \frac{f^{(n)}(0)}{n!} M_n s^{n+1} + O(s^{n+2}) \right]$$

From the admissibility condition we already have that the 0th moment $M_0 = 0$ so that the first term in the right-hand side of the above equation is zero. If we now manage to make the other moments up to $M_n$ zero as well, then the wavelet transform coefficients (s) will decay as fast as $s^{n+2}$ for a smooth signal f(t). This is known as the vanishing moments or approximation order. If a wavelet has N vanishing moments, then the approximation order of the wavelet transform is also N. The moments do not have to be exactly zero; a small value is often good enough. In fact, experimental research suggests that the number of vanishing moments required depends heavily on the application. Summarizing, the admissibility condition gave us the wave, regularity and vanishing moments gave us the fast decay, and put together they give us the wavelet (N. Fletcher and T. Rossing, 1998).

The most important properties of wavelets are the admissibility and the regularity conditions and these are the properties which gave wavelets their name. It can be shown that square integral functions ψ (t) satisfying the admissibility condition,

$$\int \frac{|\Psi(\omega)|^2}{|\omega|} d\omega < +\infty$$

This can be used to first analyze and then reconstruct a signal without loss of information. In ψ (ω) stands for the Fourier transform of ψ (t). The admissibility condition implies that the Fourier transform of ψ (t) vanishes at the zero frequency, i.e.

$$|\Psi(\omega)|^2 \Big|_{\omega=0} = 0$$

This means that wavelets must have a band-pass like spectrum (N. Fletcher and T.

Rossing, 1998). This is a very important observation, which we have used to build an efficient wavelet transform. Each signal is passed through a series of high pass filters to analyze the high frequencies, and it is passed through a series of low pass filters to analyze the low frequencies. A time compression of the wavelet by a factor of 2 will stretch the frequency spectrum of the wavelet by a factor of 2 and also shift all frequency components up by a factor of 2.

**The sub-band coding and the multi-resolution analysis**

Here we split/filtered each of the signal spectrums in two (equal) parts, a low-pass and a high-pass part. Transition band width is kept minimum across 50 Hz and the offset information is negligible thus causing no artifact in the split signal. The first high-pass part contains no information. We now have two bands. However, the low-pass part still contains some details and therefore we split it again and again, until the last low-pass part contain no information. In this way we have created an iterated filter bank. This is done by sub-band coding/pyramidal coding algorithm using Nyquist's rule (Robi Polikar). The Nyquist theorem states that perfect reconstruction of a signal is possible when the sampling frequency is greater than twice the maximum frequency of the signal being sampled or equivalently, when the Nyquist frequency (half the sample rate) exceeds the highest frequency of the signal being sampled. Why did we split the signal into several frequency bands? Energy of the signal varies with time. In this process a certain high frequency component can be located well in time while a low frequency component can be located better in frequency.  In this process of sub band coding each of the signals are divided into eight frequency sub bands of band width 7040 to 14080 rad/sec referred as 1st band or DWT 1st level, 3520 to 7040 rad/sec referred as 2nd band or DWT 2nd level, 1760 to 3520 rad/sec referred as 3rd band or DWT 3rd level, 880 to 1760 rad/sec referred as 4th band or DWT 4th level, 440 to 880 rad/sec referred as 5th band or DWT 5th level, 220 to 440 rad/sec referred as 6th band or DWT 6th level, 110 to 220 rad/sec referred as 7th band or DWT 7th level and less than 110 rad/sec referred as 8th band or DWT 8th level. From each of these sub bands number of harmonics, number of DWT coefficients and most prominent frequency (MPF) is measured. Fidelity factor Q here is ½ (Robi Polikar). By this process the scale of each signal gets doubled, reducing the uncertainty in the frequency by half.

**Observation**

DWT 1st level (7040 to 14080 rad/sec), 2nd level (3520 to 7040 rad/sec) and DWT 8th level (55 to 110 rad/sec) reveals no information. So, all information lies within 3rd to 7th level of DWT inferring that the tabla sounds are harmonically best within the range of 0.1 kHz to 3 kHz. One of the most prominent features from spectral analysis has to do with the behavior of harmonics and indeed the knowledge of locations of harmonics. We can see from figure 2 that the harmonic number increases towards mid frequency range and an expansion of the harmonic locations can be observed in band 3. It is also possible to take into account the contraction (harmonic ratio<1) or expansion (harmonic ratio>1) of harmonics in all the spectrums. Harmonic ratio is the ratio of two consecutive harmonic frequencies.  In band 5 ratio of 5th harmonic to 4th harmonic for all the strokes are less than 1, means spectral energy reduces at higher frequencies in that band. Mid frequency range dominates in band 5 while lower frequency ranges are dominating in band 6. According to Grey (Grey J M, 1977), on the distribution of the harmonics, it has been suggested that no harmonics higher than the 5th to 7th, regardless of the fundamental frequency, are resolved individually. This is true for tabla sounds and studies have shown that the upper harmonics rather than being perceived independently are heard as a group.

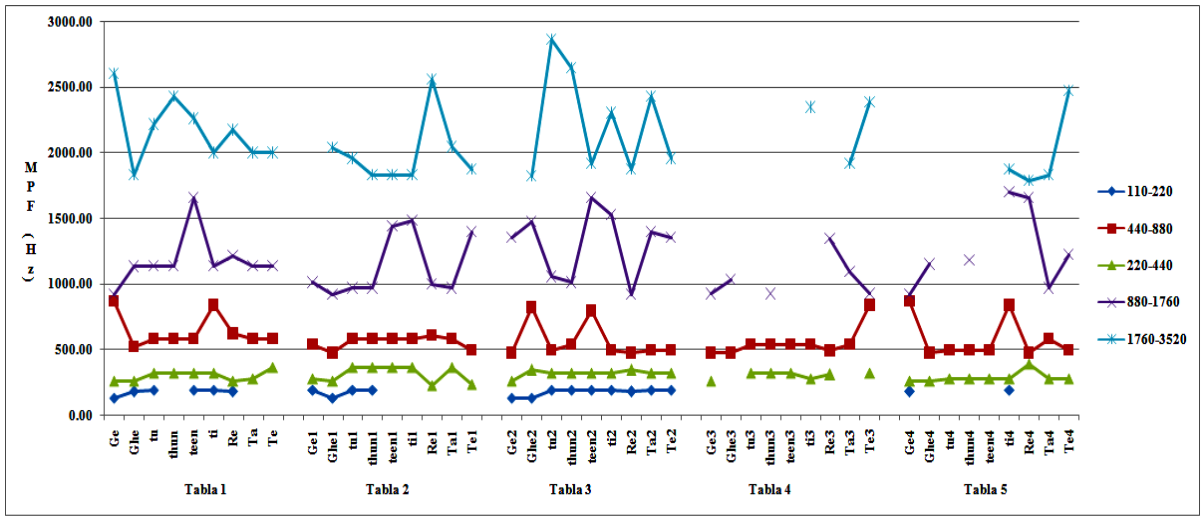

**Figure 1:** Most prominent frequency (rad/sec) at different sub-bands

Fig. 1 reveals that the MPF corresponds to the highest peak of each sub-band, is uniform during the lower frequencies up to 900Hz, but largely varies in high frequency bands. So at low frequency, all tablas sounds similar. Maximum harmonic information lies in the range 0.9 – 2.5 kHz. Although Tee and Teen, Ghe and Ge are similar strokes, a larger difference in MPF occurs for all the first three tablas. Such difference is due to the damping effect among the pair of strokes and is visible in 440 to 1760 Hz. But 4th and 5th tablas give no information during these ranges. Such differences give rise to the fact that the production of harmonics is different for all tablas due to difference in their structure i.e. the resonant chamber.

Stroke Ta/Na (where the position of stroke and damping are different) of all tablas shows that the attack peak forms late compared to all other strokes. Attack peak of all strokes of 1st tabla delays to reach (found in the 6th of its band structure). Attack peak of all strokes of 2nd tabla immediately reaches (found in the 7th of its band structure). So the players of 1st and 2nd tablas made all the strokes consistently. No such consistency is found in achieving attack peak for the rest of the three tablas because of their difference in stroke production. So the feature attack peak is an important cue to assess the learners about the accuracy of stroke both in terms of stroke intensity and also position. Attack peak also depend upon the stroke type and position viz. stroke Thun executes by striking on the centre circle with index, middle, ring and little fingers together and by quickly releasing, reaches the attack peak immediately while stroke Ta/Na executes by lightly pressing the ring finger down in order to mute the sound while index finger strikes the edge, reaches the attack peak delayed. It is also found that attack peak reaches faster in free strokes than damped strokes.

When a membrane stretched over a resonating body is struck, there is generally no clear sense of pitch because the sound produced is rich in inharmonic overtones. When properly applied, the syahi (outside the central part) causes the alignment of some of the inharmonic partials, giving the dayana clear sense of pitch if struck correctly (N Fletcher and T Rossing, 1998).

Also it is observed a shorter decay for Re, Te, Tu and Na while longer decay for Ghe, Ge, Tee and Teen. MPF of damped strokes show lesser time of occurrence compared to free strokes. Calculating the scale (frequency) and time of all the harmonics, it is observed that lower harmonics in the attack portion are always lagging. It is also observed that the higher energy bands are of shorter duration than the lower energy bands. In comparison to other strokes, Teen has longer duration of energy in all the bands. This proves that tabla produces different sounds with definite harmonic properties with the variation of stroke production and position.

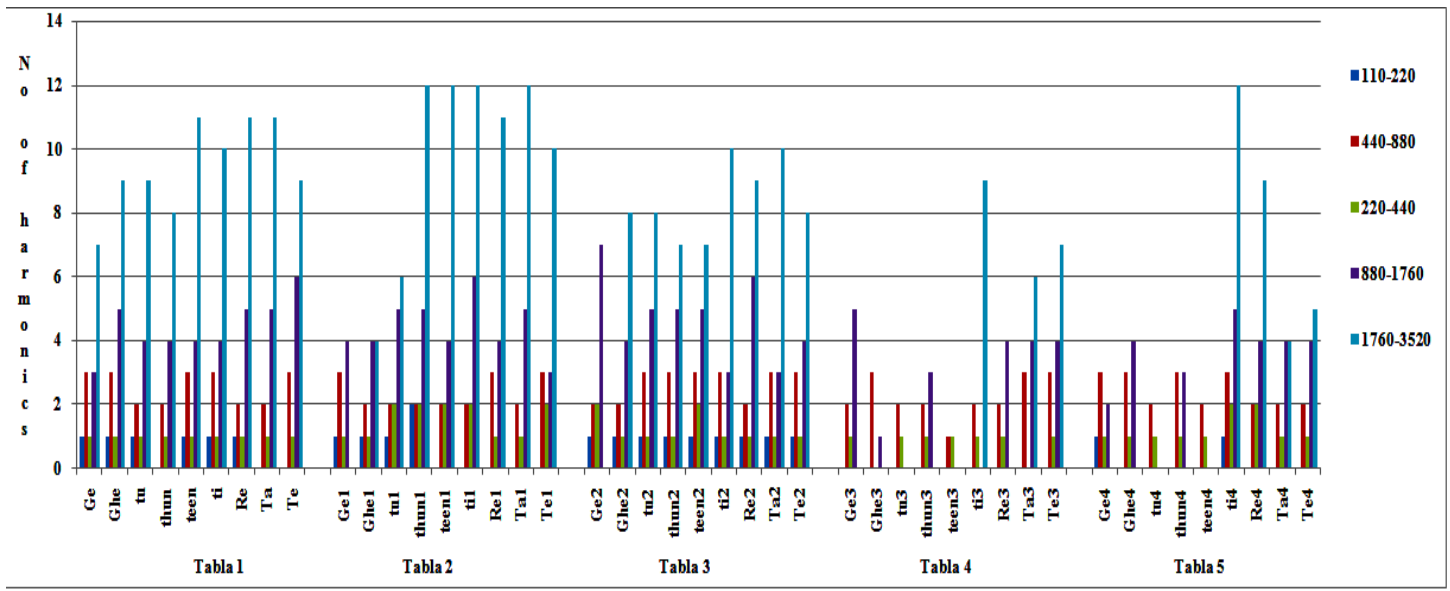

**Figure 2:** Number of harmonics at different sub-bands

From fig 2, Damped strokes have larger number of harmonics at high frequency ranges (band 3 and 4) for all the tablas compared to free strokes. Comparing the number of harmonics at higher frequency range we can categorize the stroke of tablas. For the 1st, 2nd and 3rd tablas (all are smaller in size), all strokes produce harmonics uniformly and equally in bands 5 and 6 but at higher frequency bands they show larger number of harmonics. For the 1st, 2nd and 3rd tablas, harmonics are present within band 3 to band 6. 4th and 5th tablas (both are bigger in size) produces very less number of harmonics for the free strokes and they are found within band 4 and band 6 but the damped strokes produces harmonics within band 3 and band 6. So it is observed that tablas with smaller diameter produce more harmonics than tablas of larger diameter.

**Table 1:** Standard deviation and mean of tabla strokes at different sub-bands

|  | 110-220 | | 220-440 | | 440-880 | | 880-1760 | | 1760-3520 | |
| --- | --- | --- | --- | --- | --- | --- | --- | --- | --- | --- |
|  | Stdev | Mean | Stdev | Mean | Stdev | Mean | Stdev | Mean | Stdev | Mean |
| Tabla 1 | 24.341 | 178.732 | 36.927 | 302.998 | 124.346 | 640.864 | 195.905 | 1181.595 | 242.300 | 2170.102 |
| Tabla 2 | 31.433 | 178.149 | 61.536 | 315.576 | 45.133 | 558.249 | 236.912 | 1130.017 | 245.908 | 1996.191 |
| Tabla 3 | 27.200 | 178.310 | 23.820 | 321.220 | 139.530 | 565.568 | 253.916 | 1307.372 | 393.874 | 2228.098 |
| Tabla 4 |  |  | 24.889 | 306.703 | 111.122 | 552.938 | 165.671 | 1043.175 | 261.962 | 2217.92 |
| Tabla 5 |  |  | 40.109 | 288.843 | 159.109 | 580.582 | 309.175 | 1258.977 | 324.960 | 1991.757 |

Again table 1 shows the standard deviation and average frequency of harmonics. It is observed that at low frequency range both the standard deviation and average frequency remains same but at higher frequency bands these values get scattered. This observation is statistically confirmed in the later section. Hence the randomness of energy of the partial increases at higher frequency bands and uniqueness of table sound lies in the higher energy. From the data it is observed that 4th tabla is totally different compared to others. Such thing occurs due to differences in structure of the tabla as well as the intensity of strokes made by the player.

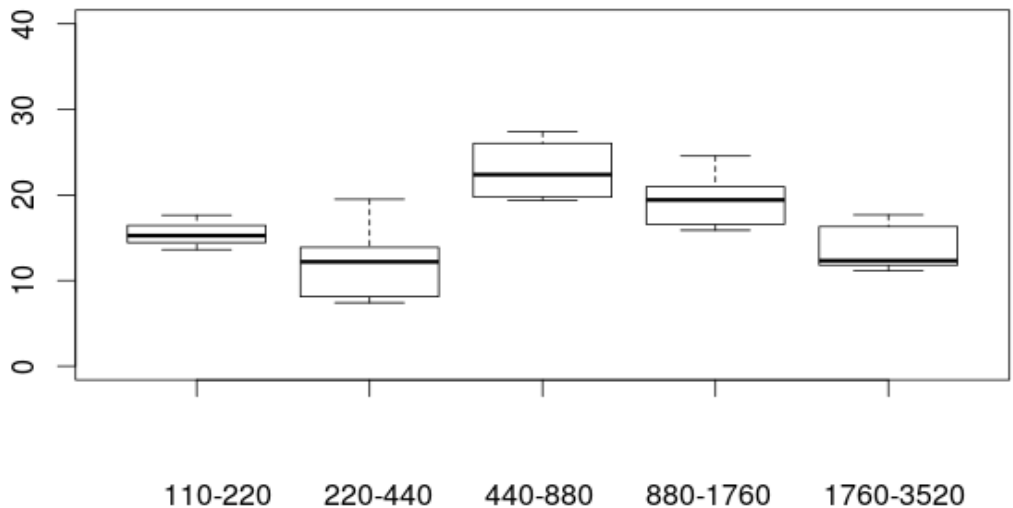

## Boxplots of Coefficient of Variations of Different Sub Bands

Now statistically, we will look into the patterns depicted by harmonics of different sub bands and the tablas. This analysis will be done in three different ways, i.e. we will look into the means, standard deviations and coefficient of variation from Table 1 and try to study that whether there is any significant difference in the means, standard deviation and coefficient of variation of harmonics of the tablas and sub bands. The main idea behind considering coefficient of variation is that it incorporates both mean and standard deviation and gives a clear understanding of the underlying pattern. These comparisons will be done using One-Way ANOVA and if there is any significant difference, we will go for detailed pair wise comparison using Tukey HSD method. To complement our findings, we will be using Welch Robust Test of Equality of means. We will use SPSS and R for all the analyses.

## Coefficient of Variation of Harmonics of Different Sub Bands

| | N | Mean | Std. Deviation | Std. Error | 95% Confidence Interval for Mean | | Minimum | Maximum |
|---|---|---|---|---|---|---|---|---|
| | | | | | Lower Bound | Upper Bound | | |
| 1.00 | 3 | 15.5057552 | 2.02449572 | 1.16884315 | 10.4766290 | 20.5348814 | 13.61871 | 17.64422 |
| 2.00 | 5 | 12.2206753 | 4.89140967 | 2.18750490 | 6.1471880 | 18.2941626 | 7.41548 | 19.49958 |
| 3.00 | 5 | 19.9320239 | 7.39877645 | 3.30883342 | 10.7452296 | 29.1188183 | 8.08474 | 27.40509 |
| 4.00 | 5 | 19.4811952 | 3.51260241 | 1.57088355 | 15.1197232 | 23.8426671 | 15.88142 | 24.55764 |
| 5.00 | 5 | 13.8576447 | 2.93411206 | 1.31217480 | 10.2144634 | 17.5008260 | 11.16537 | 17.67759 |
| Total | 23 | 16.2597809 | 5.38559134 | 1.12297342 | 13.9308766 | 18.5886853 | 7.41548 | 27.40509 |

**Descriptives**

The above table gives all the descriptive details.

The test of Homogeneity of Variance shows whether the variances of harmonics are more or less same among different groups. The p-value is greater than 0.05 (p value is 0.487), hence we conclude that by levene's test (levene's statistics: 0.896 with degrees of freedom (4,18)), the group variances are homogeneous in nature. The

ANOVA table shows that, the p-value is greater than 0.05 (p-value 0.074 and and F(4,18)=2.561). Hence, the coefficient of variations of harmonics of the different sub bands doesn't differ significantly. The Welch Robust Test of Equality of Means also shows the same, i.e. as the p-value is highly greater than 0.05 (p-value 0.126), we conclude that the group means of harmonics of the coefficient of variations of the sub bands do not differ significantly.

## Coefficient of Variation of Different Tablas

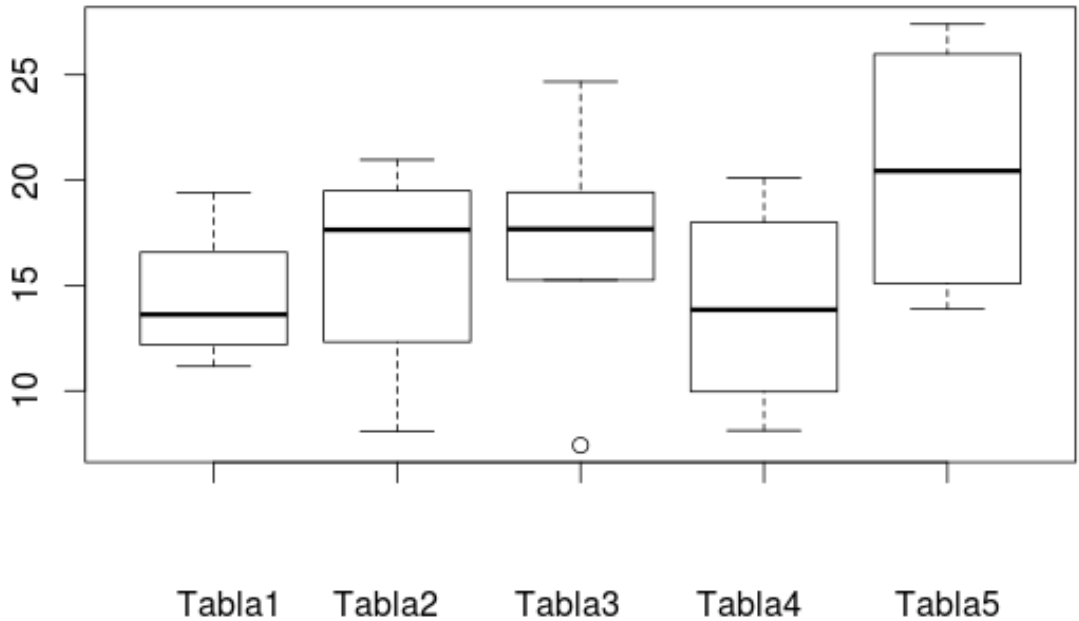

| | | | | | | | | |
|---|---|---|---|---|---|---|---|---|
| **Descriptives** ||||||||| 
| Tabla cv ||||||||| 
| | N | Mean | Std. Deviation | Std. Error | 95% Confidence Interval for Mean || Minimum | Maximum |
| | | | | | Lower Bound | Upper Bound | | |
| 1 | 5 | 14.59078 | 3.375191 | 1.509431 | 10.39993 | 18.78163 | 11.165 | 19.403 |
| 2 | 5 | 15.70254 | 5.370402 | 2.401717 | 9.03431 | 22.37077 | 8.085 | 20.965 |
| 3 | 5 | 16.88802 | 6.324819 | 2.828545 | 9.03472 | 24.74132 | 7.416 | 24.671 |
| 4 | 4 | 13.97605 | 5.168171 | 2.584085 | 5.75234 | 22.19976 | 8.115 | 20.097 |
| 5 | 4 | 20.54100 | 6.465138 | 3.232569 | 10.25352 | 30.82848 | 13.886 | 27.405 |
| Total | 23 | 16.25978 | 5.385590 | 1.122973 | 13.93087 | 18.58868 | 7.416 | 27.405 |

From the boxplot itself we observe a definite pattern. Tabla 4 has lowest coefficient of variation in harmonics. The above table gives all the descriptive details for the four tablas for coefficient of variations of harmonics. We will first undertake the test of Homogeneity of Variance to find whether the variances of coefficient of variations of harmonics are more or less same among different tablas. As the p value is greater than 0.05 (p value: 0.592, levene's statistics: 0.716, with degrees of freedom (4,18)),we conclude by levene's test, that the group variances of coefficient of variations of different tablas are homogeneous. Now we go for one way ANOVA to understand whether there is any significant difference in the means of the coefficient of variations of harmonics of different tablas.

The ANOVA table shows that, the p-value is highly greater than 0.05 (p-value 0.456 and $F(4,18)=0.954$). Hence, the coefficient of variations of the different tablas for harmonics does not differ significantly. The Welch robust test of equality of Means also shows the same, i.e. as the p-value is highly greater than 0.05 (p-value 0.621), the group means of the coefficient of variations of the tablas for harmonics do not differ significantly.

## Means of Different Sub Bands

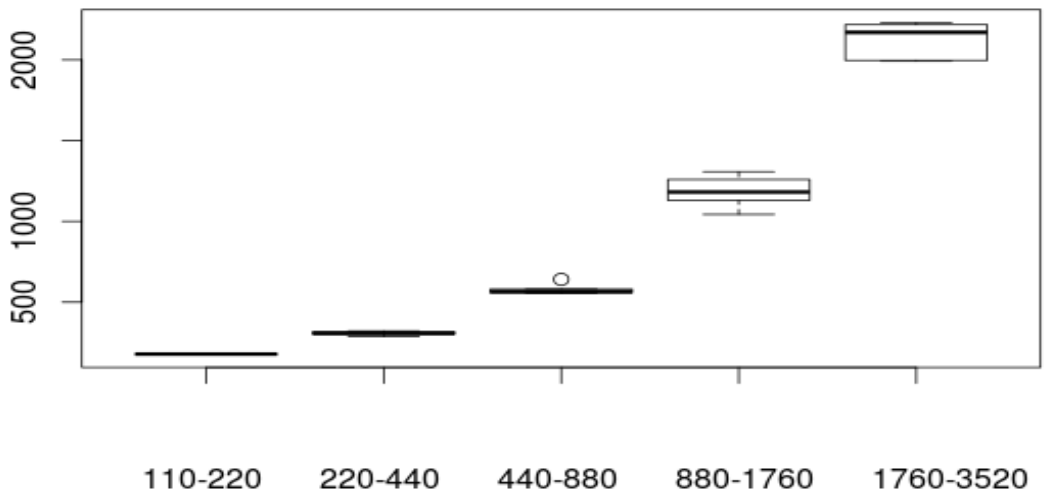

| | | | | | | | | |
|---|---|---|---|---|---|---|---|---|
| **Descriptives** | | | | | | | | |
| | N | Mean | Std. Deviation | Std. Error | 95% Confidence Interval for Mean | | Minimum | Maximum |
| | | | | | Lower Bound | Upper Bound | | |
| 1.00 | 3 | 178.3970000 | .30107972 | .17382846 | 177.6490765 | 179.1449235 | 178.14900 | 178.73200 |
| 2.00 | 5 | 307.0680000 | 12.46509043 | 5.57455791 | 291.5905460 | 322.5454540 | 288.84300 | 321.22000 |
| 3.00 | 5 | 579.6402000 | 35.76960445 | 15.99665341 | 535.2263699 | 624.0540301 | 552.93800 | 640.86400 |
| 4.00 | 5 | 1184.2272000 | 104.39536838 | 46.68702804 | 1054.6032295 | 1313.8511705 | 1043.17500 | 1307.37200 |
| 5.00 | 5 | 2120.8136000 | 117.85090179 | 52.70452552 | 1974.4823781 | 2267.1448219 | 1991.75700 | 2228.09800 |
| Total | 23 | 934.5189565 | 731.32705712 | 152.49223280 | 618.2694219 | 1250.7684912 | 178.14900 | 2228.09800 |

From the box plot itself it is clear that the mean of harmonics of different sub bands differ highly. Also it seems that the variability in the means do increase with the advancement in the sub bands. The above table gives all the descriptive details in this regard. As the p value is less than 0.05 (p-value is 0.001 and levene statistic 8.774, degrees of freedom (4,18)), we conclude by levene's test, the group variances of means of harmonics of different Sub-bands are not homogeneous in nature.

Now to understand whether there is any significant difference between the means of the sub-bands, we undertake one way ANOVA. The p-value for the ANOVA table is less than 0.05, (p-value 0.000, $F(4,18)=500.313$). Hence, we conclude that, the sub-band means of harmonics do differ significantly; same kind of scenario has been depicted in the robust test also. It shows that the p-value is less than 0.001; hence the sub band means of harmonics do differ significantly.

Now as we have obtained, that the means do differ significantly, we would go for post-hoc tests by TUKEY HSD to find which sub band means are same and which ones differ among them. Sub-band 1 and Sub- band 2 doesn't differ significantly (p-value 0.188), but the other sub bands do differ significantly among themselves.

| **Multiple Comparisons** | | | | | | |
|---|---|---|---|---|---|---|
| Tukey HSD | | | | | | |
| (I) indicator | (J) indicator | Mean Difference (I-J) | Std. Error | Sig. | 95% Confidence Interval | |
| | | | | | Lower Bound | Upper Bound |
| 1.00 | 1.00 | | | | | |
| | 2.00 | -128.67100000 | 55.74770298 | .188 | -297.2406692 | 39.8986692 |
| | 3.00 | -401.24320000* | 55.74770298 | .000 | -569.8128692 | -232.6735308 |
| | 4.00 | -1005.83020000* | 55.74770298 | .000 | -1174.3998692 | -837.2605308 |
| | 5.00 | -1942.41660000* | 55.74770298 | .000 | -2110.9862692 | -1773.8469308 |
| 2.00 | 1.00 | 128.67100000 | 55.74770298 | .188 | -39.8986692 | 297.2406692 |
| | 2.00 | | | | | |
| | 3.00 | -272.57220000* | 48.27892699 | .000 | -418.5578159 | -126.5865841 |
| | 4.00 | -877.15920000* | 48.27892699 | .000 | -1023.1448159 | -731.1735841 |
| | 5.00 | -1813.74560000* | 48.27892699 | .000 | -1959.7312159 | -1667.7599841 |
| 3.00 | 1.00 | 401.24320000* | 55.74770298 | .000 | 232.6735308 | 569.8128692 |
| | 2.00 | 272.57220000* | 48.27892699 | .000 | 126.5865841 | 418.5578159 |
| | 3.00 | | | | | |
| | 4.00 | -604.58700000* | 48.27892699 | .000 | -750.5726159 | -458.6013841 |
| | 5.00 | -1541.17340000* | 48.27892699 | .000 | -1687.1590159 | -1395.1877841 |
| 4.00 | 1.00 | 1005.83020000* | 55.74770298 | .000 | 837.2605308 | 1174.3998692 |
| | 2.00 | 877.15920000* | 48.27892699 | .000 | 731.1735841 | 1023.1448159 |
| | 3.00 | 604.58700000* | 48.27892699 | .000 | 458.6013841 | 750.5726159 |
| | 4.00 | | | | | |
| | 5.00 | -936.58640000* | 48.27892699 | .000 | -1082.5720159 | -790.6007841 |
| 5.00 | 1.00 | 1942.41660000* | 55.74770298 | .000 | 1773.8469308 | 2110.9862692 |
| | 2.00 | 1813.74560000* | 48.27892699 | .000 | 1667.7599841 | 1959.7312159 |
| | 3.00 | 1541.17340000* | 48.27892699 | .000 | 1395.1877841 | 1687.1590159 |
| | 4.00 | 936.58640000* | 48.27892699 | .000 | 790.6007841 | 1082.5720159 |
| | 5.00 | | | | | |
| *. The mean difference is significant at the 0.05 level. | | | | | | |

# Standard Deviations for Different Sub-bands

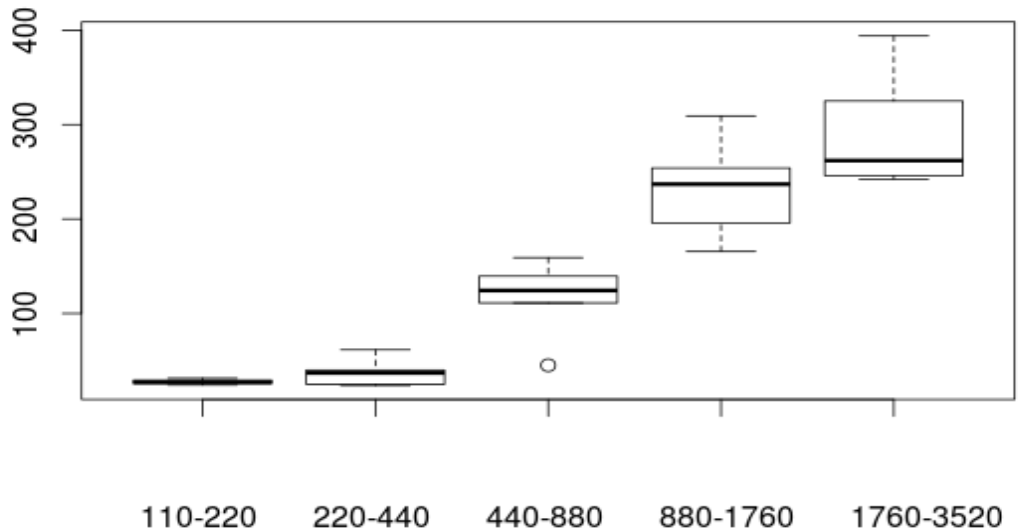

| Descriptives | | | | | | | | |
|---|---|---|---|---|---|---|---|---|
| **Sdsubband** | | | | | | | | |
| | N | Mean | Std. Deviation | Std. Error | 95% Confidence Interval for Mean | | Minimum | Maximum |
| | | | | | Lower Bound | Upper Bound | | |
| 1.00 | 3 | 27.6580000 | 3.56811421 | 2.06005170 | 18.7943129 | 36.5216871 | 24.34100 | 31.43300 |
| 2.00 | 5 | 37.4562000 | 15.25643008 | 6.82288295 | 18.5128400 | 56.3995600 | 23.82000 | 61.53600 |
| 3.00 | 5 | 115.8480000 | 43.38062226 | 19.40040406 | 61.9838431 | 169.7121569 | 45.13300 | 159.10900 |
| 4.00 | 5 | 232.3158000 | 55.14142282 | 24.65999396 | 163.8486805 | 300.7829195 | 165.67100 | 309.17500 |
| 5.00 | 5 | 293.8008000 | 65.08706281 | 29.10781938 | 212.9845374 | 374.6170626 | 242.30000 | 393.87400 |
| Total | 23 | 151.3077391 | 114.59218519 | 23.89412235 | 101.7543623 | 200.8611160 | 23.82000 | 393.87400 |

The above table gives all the descriptive details for the standard deviations (sd) of harmonics of different sub-bands. The box plot shows an important trend, that the variability in the standard deviation values of the harmonics of the sub bands increases with the advancement the the sub bands. Now we need to test this claim statistically.

Like previous analyses, we first start with the test of Homogeneity of Variance. As the p value is less than 0.05 (p-value is 0.037, levene's statistic is 3.222, degrees of freedom is (4,18)), we conclude that by levene's test, the group variances are not homogeneous in nature. Now we go for one-way ANOVA for more detailed analysis. The p-value for the anova table is less than 0.05, (p-value 0.000, F(4,18)=30.082), hence, we conclude that, the sub band sds of harmonics do differ significantly, same kind of scenario has been depicted in the robust test also. It shows that the p-value is 0.000, hence the sub band sds of the harmonics do differ significantly. Now as we have obtained, that the sds of harmonics of sub bands do differ significantly, we would go for post-hoc tests by TUKEY HSD to find which sub band sds are same and which differ among themselves.

Both Sub-band 1 and sub- band 2 differs significantly with sub-band 4 and sub-band 5 (p-value is less than 0.001 for all the cases). Sub-band 3 doesn't differs significantly from sub-band 1 (p-value is 0.104), whereas both sub band 4 and sub-band 5 differ significantly from sub-band 1 and sub-band 2 significantly (p-value less than 0.05). This same pattern is also observed in the box plots also.

| \multicolumn{7}{c}{**Multiple Comparisons**} |
| :--- | :--- | :--- | :--- | :--- | :--- | :--- |
| Tukey HSD | | | | | | |
| (I) indicator | (J) indicator | Mean Difference (I-J) | Std. Error | Sig. | 95% Confidence Interval | |
| | | | | | Lower Bound | Upper Bound |
| 1.00 | 1.00 | | | | | |
| | 2.00 | -9.79820000 | 33.37402853 | .998 | -110.7144468 | 91.1180468 |
| | 3.00 | -88.19000000 | 33.37402853 | .104 | -189.1062468 | 12.7262468 |
| | 4.00 | -204.65780000* | 33.37402853 | .000 | -305.5740468 | -103.7415532 |
| | 5.00 | -266.14280000* | 33.37402853 | .000 | -367.0590468 | -165.2265532 |
| 2.00 | 1.00 | 9.79820000 | 33.37402853 | .998 | -91.1180468 | 110.7144468 |
| | 2.00 | | | | | |
| | 3.00 | -78.39180000 | 28.90275653 | .091 | -165.7878334 | 9.0042334 |
| | 4.00 | -194.85960000* | 28.90275653 | .000 | -282.2556334 | -107.4635666 |
| | 5.00 | -256.34460000* | 28.90275653 | .000 | -343.7406334 | -168.9485666 |
| 3.00 | 1.00 | 88.19000000 | 33.37402853 | .104 | -12.7262468 | 189.1062468 |
| | 2.00 | 78.39180000 | 28.90275653 | .091 | -9.0042334 | 165.7878334 |
| | 3.00 | | | | | |
| | 4.00 | -116.46780000* | 28.90275653 | .006 | -203.8638334 | -29.0717666 |
| | 5.00 | -177.95280000* | 28.90275653 | .000 | -265.3488334 | -90.5567666 |
| 4.00 | 1.00 | 204.65780000* | 33.37402853 | .000 | 103.7415532 | 305.5740468 |
| | 2.00 | 194.85960000* | 28.90275653 | .000 | 107.4635666 | 282.2556334 |
| | 3.00 | 116.46780000* | 28.90275653 | .006 | 29.0717666 | 203.8638334 |
| | 4.00 | | | | | |
| | 5.00 | -61.48500000 | 28.90275653 | .252 | -148.8810334 | 25.9110334 |
| 5.00 | 1.00 | 266.14280000* | 33.37402853 | .000 | 165.2265532 | 367.0590468 |
| | 2.00 | 256.34460000* | 28.90275653 | .000 | 168.9485666 | 343.7406334 |
| | 3.00 | 177.95280000* | 28.90275653 | .000 | 90.5567666 | 265.3488334 |
| | 4.00 | 61.48500000 | 28.90275653 | .252 | -25.9110334 | 148.8810334 |
| | 5.00 | | | | | |
| *. The mean difference is significant at the 0.05 level. | | | | | | |

**Conclusion**

With the help of wavelet tools and the statistical analyses done so far, we can precisely distinguish individual strokes which may lead to the following conclusions

(i) Categorization of tabla viz. (a) tabla 1 differs from other four tablas; (b) tablas with smaller diameter produce more harmonics than tablas of larger diameter significantly; (c) stroke at the edge (3rd circle) produces weak resonance in the cavity of tabla and hence produce low energy sound while stroke at the 2nd circle produces strong resonance in the cavity of tabla and hence produce high energy sound, (d) tabla sounds are harmonically best within the range of 0.1 kHz to 3 kHz.

(ii) Categorization of tabla strokes viz. (a) damped strokes are more powerful (energetic), higher irregularity in its harmonics and also possess larger number of harmonics than free strokes; (b) tabla strokes have weak fundamental, (c) attack peak reaches faster in free strokes than the damped strokes,

(d) time to reach MPF is less for damped strokes compared to free strokes. Hence this percussion instrument is different from others.

**References**

- B M Banerjee and D Nag, "*The Acoustical Character of Sounds from Indian Twin Drums*", ACUSTICA, Vol.75, 206-208, (1991)
- B S Ramakrishna "*Modes of Vibration of the Indian Drum Dugga or left hand Thabala*", J. Acoust. Soc. Am. Vol.29, 234-238, (1957)
- C V Raman "*The Indian Musical Drums*", Proc. Indian Acad. Sc., 1A, pp. 179-188, (1934)
- C V Raman and Sivakali Kumar "*Musical drums with harmonic overtones*", Nature, Vol.104, 500, (1920)
- C. Valens, "*A Really Friendly Guide to Wavelets (C)*", (1999-2004), wavelets@polyvalens.com
- David Courtney "*Psychoacoustics of the Musical pitch of Tabla*" Journal of SanGeet Research Academy, Calcutta, India, Vol.13, No 1, (1999)
- Grey J. M. "*Multidimensional Perceptual Scaling of Musical Timbre*", Journal of the Acoustical Society of America Vol. 61, 1270 – 1277. (1977)
- K N Rao "Theory of the Indian Musical Drums", Proc. Indian Acad. Sci. Vol.7A, 75-84, (1938)
- N. Fletcher and T. Rossing. "*The physics of musical instruments*", Springer, New York, (1998).
- Paragchordia, "*Segmentation and recognition of tabla strokes*", CCRMA, Stanford university, Pchordia@ccrma.stanford.edu
- R N Ghosh "*Note on Musical Drums*", Phys. Review, Vol.20, 526-527, (1922)
- RobiPolikar,'*The Wavelet Tutorial*', Rowan University, http://users.rowan.edu/polikar/wavelets/wtutorial.html
- T Sarojini and A Rahman "*Variational Method for the Vibrations of Indian Drums*" J. Acoust. Soc. Am. Vol.30, 191-196, (1958)